\documentclass[aps,prl,twocolumn,superscriptaddress]{revtex4}
 \usepackage{graphicx}
\usepackage{afterpage} \usepackage{dcolumn} \usepackage{bm}
\usepackage{color}

\begin{document}

\preprint{APS/123-QED}

\title{Spin resonance in EuTiO$_3$ probed by time-domain GHz ellipsometry}

\author{J.L.M.~van~Mechelen}
\affiliation{D\'epartement de Physique de la Mati\`ere
Condens\'ee, Universit\'e de Gen\`eve, Gen\`eve, Switzerland.}

\author{D.~van~der~Marel}
\affiliation{D\'epartement de Physique de la Mati\`ere
Condens\'ee, Universit\'e de Gen\`eve, Gen\`eve, Switzerland.}

\author{I.~Crassee}
\affiliation{D\'epartement de Physique de la Mati\`ere
Condens\'ee, Universit\'e de Gen\`eve, Gen\`eve, Switzerland.}

\author{T.~Kolodiazhnyi}
\affiliation{National Institute of Materials Science, Tsukuba, Ibaraki, Japan.}

\date{\today}
\begin{abstract}We show an example of a purely magnetic spin resonance in EuTiO$_3$ and the resulting new record high Faraday rotation of 590 deg/mm at 1.6\,T for 1 cm wavelengths probed by a novel technique of magneto-optical GHz time-domain ellipsometry. From our transmission measurements of linear polarized light we map out the complex index of refraction $n=\sqrt{\epsilon\mu}$ in the GHz to THz range. We observe a strong resonant absorption by magnetic dipole transitions involving the Zeeman split $S=7/2$ magnetic energy levels of the Eu$^{2+}$ ions, which causes a very large dichroism for circular polarized radiation.
\end{abstract}

\maketitle

The magnetic state of solids can be measured and manipulated through the interaction with photons. The interaction of the electric field component of electromagnetic radiation with the orbital motion of electrons is described by the dielectric permeability $\epsilon_{\pm}(\omega)$, where the suffix indicates left handed (LCP, $-$) and right handed circular polarization (RCP, $+$). Interaction with the electron spin occurs through the magnetic field component and is described by the magnetic susceptibility $\mu_{\pm}(\omega)$. Photons interact differently with magnetically polarized matter depending on whether their angular momentum ($\pm\, \hbar$) is parallel or antiparallel to the magnetic polarization. This so-called circular dichroism transforms incident linear polarized light (which is a superposition of equal amounts of LCP and RCP) to elliptical after transmission (Faraday effect) or reflection (Kerr effect).
The magnetic field component of the electromagnetic field in the THz range is the relevant component which interacts with the local magnetic moments in EuTiO$_3$, because the 4f$^7$ ground state of the Eu$^{2+}$ ions has only spin- and no orbital component. Photons in the THz range, when absorbed by a 4f$^7$ moment, transfer their angular momentum with the matching chirality to the magnetic ion.

Bulk EuTiO$_3$ has the same perovskite crystal structure as room temperature SrTiO$_3$. It is magnetic due to the seven parallel spins on the Eu site and has thus $S=7/2$ and $L=0$ and a Land\'e $g$-factor of 2. Neutron diffraction at low temperatures has shown that EuTiO$_3$ orders antiferromagnetically below $T_N=5.5$~K in a G-type structure~\cite{mcguire}. The rare-earth perovskite phase diagram shows EuTiO$_3$ to be on the borderline between AFM and FM ordering~\cite{komarek}. At 4.5\,K application of a small external magnetic field of about 0.2 T provokes a spin-flop transition and above 0.7\,T the magnetization rapidly saturates~\cite{shvartsman10}. EuTiO$_3$ can be made FM by applying a biaxial tensile strain of 1.1\%~\cite{lee10}. Bulk EuTiO$_3$ is a band insulator and is transparent in the THz range.

\begin{figure}
\centering
\includegraphics[width=\columnwidth]{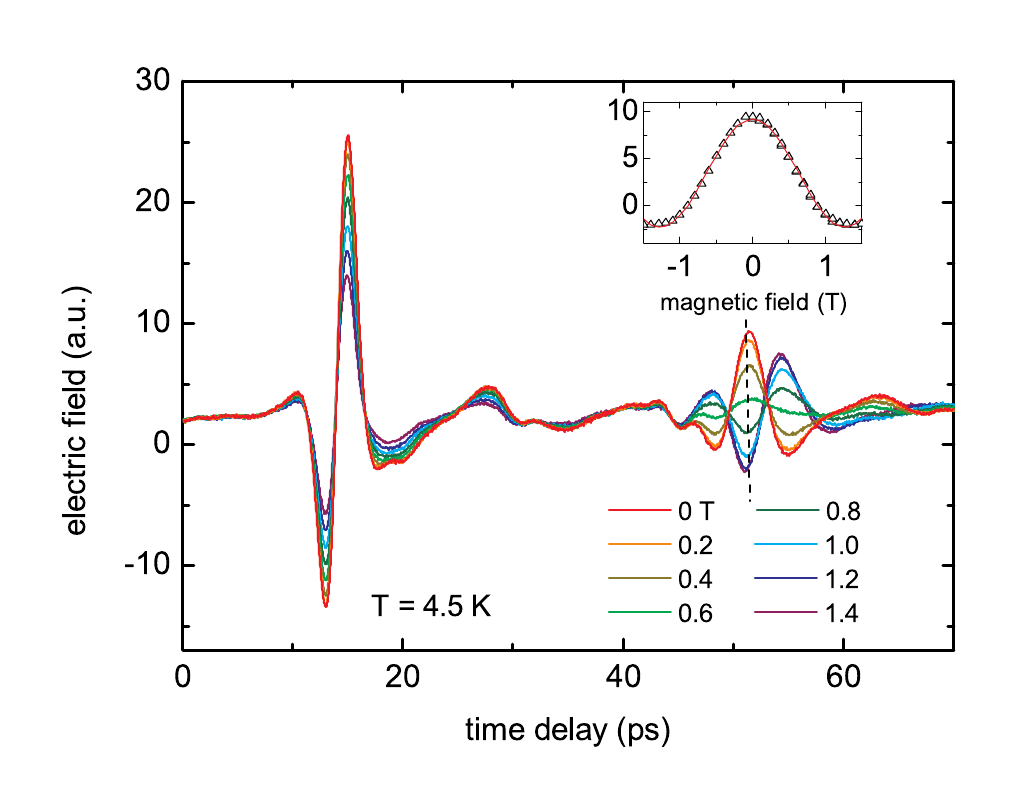}
\caption{Transmitted electric field vs.\ time delay through a 315 $\mu$m thick EuTiO$_3$ slab for selected values of the applied magnetic field at 4.5 K. The incident radiation is linearly polarized and after transmission is recorded by a linearly polarized detector. The inset shows the electric field as a function of the magnetic field at 51 ps; the red curve is a cosine function.}
\label{fig1}
\end{figure}

Here we present a spectroscopic study of the spin dynamics of insulating EuTiO$_3$ probed in the GHz to THz range simultaneously for LCP and RCP. In prior work on AFM and FM spin resonances in various oxides~\cite{snoek47,rado50,mihaly07,mihaly04,talbayev04,langner09} both spin and orbital degrees of freedom determine the resonance spectrum. In the present study we directly probe the $4f$ orbitals of EuTiO$_3$ which allows us to investigate a purely magnetic spin resonance. Highly dense (97-98\,\%) polycrystalline EuTiO$_3$ pellets with the SrTiO$_3$ structure were synthesized by spark plasma sintering as described in Ref.~\cite{vanmechelen_thesis}. Optical measurements are performed on circular samples with a thickness of 265  $\mu$m and 315 $\mu$m and a diameter of 10 mm, resulting from different fabrication batches. Optical polishing of both samples resulted in a black optical appearance with almost single crystal like shiny surfaces.

The transmitted electric field $E$ of a linearly polarized GHz-THz pulse was measured in the time-domain at 4.5\,K (see Fig.~\ref{fig1}). The emitting and receiving antennas of the spectrometer (Teraview Ltd., Cambridge, UK) are linearly polarized. The frequency range has been extended from 60 GHz down to 25 GHz by choosing an appropriate Austin switch emitter and receiver. As a function of time delay, Fig.~\ref{fig1} shows a pulse which corresponds to the directly transmitted beam followed by a pulse which exits the material after two internal reflections. Application of a magnetic field $H$, parallel to the propagation of the light, has the effect of initially decreasing the amplitudes of both pulses. Demagnetization effects are negligible for this sample/field geometry. For the second pulse the influence of $H$ is approximately three times stronger than for the first one. As a function of increasing field, $E(t)$ passes through zero and starts to grow again with opposite sign. The inset of Fig.~\ref{fig1} clearly demonstrates the periodic cosine dependence on $H$. This behavior can be understood if we assume that the photons, which are linearly polarized, pick up a Faraday rotation $\theta_F$ proportional to $H$ after passing through the sample once. Taking into account that the detecting and emitting antenna are aligned parallel to each other, the detected field amplitude of the $j$th pulse (which has traversed the sample $1+2j$ times due to multiple internal reflection) is expected to be proportional to $\cos[(1+2j)\theta_F]$. This is indeed the experimentally observed behavior displayed in the inset of Fig.~\ref{fig1}.

\begin{figure}
\centering
\includegraphics[width=\columnwidth]{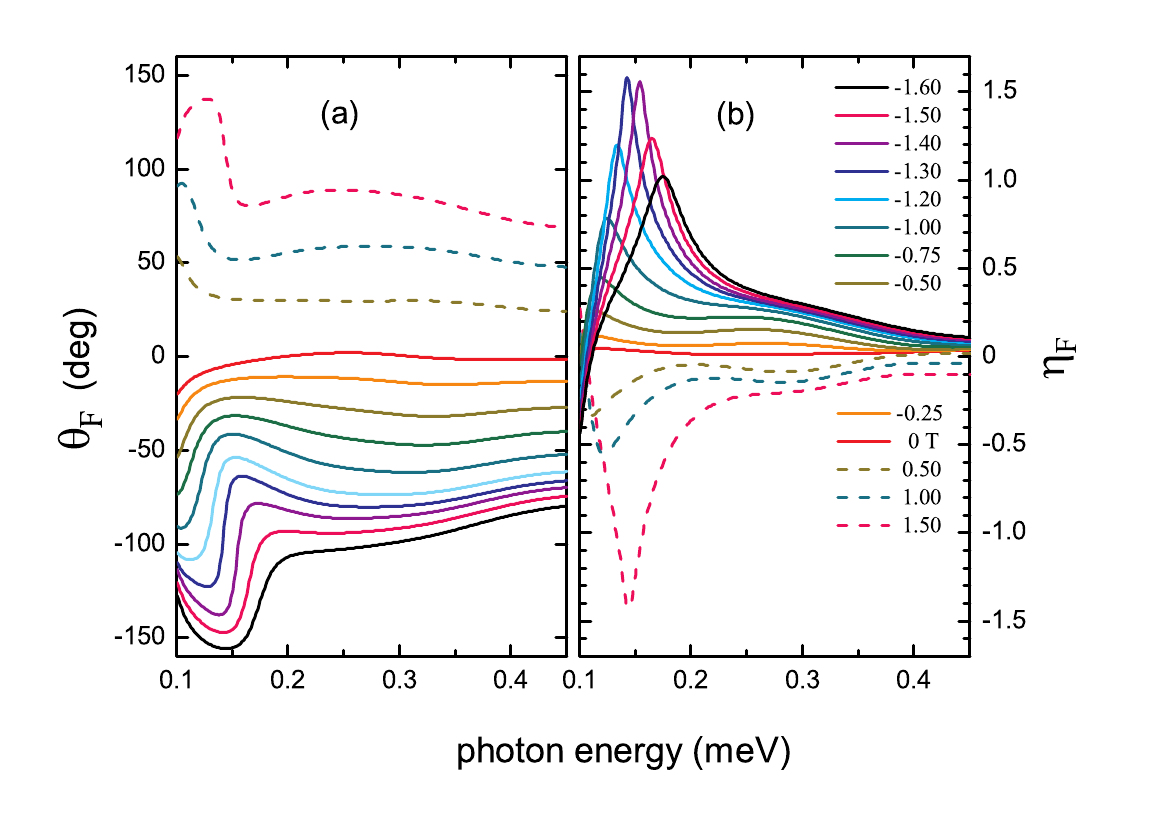}
\caption{ The (a) real and (b) imaginary part of the complex Faraday angle $\Theta_{\text{F}}=\theta_{\text{F}}+i\eta_{\text{F}}$ of the first transmitted pulse as a function of photon energy for selected magnetic fields at 4.5 K.}
\label{fig2}
\end{figure}

In order to determine the complex index of refraction $n_\pm(\omega)=\sqrt{\mu_{\pm}(\omega)\epsilon_{\pm}(\omega)}$ for both chiralities, we measured the change of polarization state of the light after transmission using a fixed linear wire-grid polarizer in front of the sample and an identical rotatable polarizer behind the sample. In this way, the time dependent electric field strength $E(t)$ is obtained for analyzer angles $\alpha$ up to over $2\pi$ radians. Making use of the Jones matrix formalism, the electric field at each time $t$ has been fitted to the relationship
\begin{equation}
E(\alpha)|_t =E_{xx}(t)\cos^2\alpha-E_{xy}(t)\cos\alpha\sin\alpha.
\end{equation}
This gives the transmitted electric fields $E_{xx}$ and $E_{xy}$ in the time-domain. Fourier transformation of both quantities provides the complex electric fields $E_\pm(\omega)$ for each chirality. For the Fourier transformation, the signal in the time domain $0<t<30$\,ps was used. This corresponds to the signal which has passed through the sample without multiple reflections and limits the frequency resolution to 0.1 meV. Consequently the frequency domain spectra are smooth on the same energy scale. We observe that $E_+(\omega)$ and $E_-(\omega)$ are significantly different, as will be discussed later. This causes the incoming polarization to change from linear to elliptical, in agreement with the result from Fig.~\ref{fig1}. The ellipse is characterized by the complex Faraday angle $\Theta_{\text{F}}=\theta_{\text{F}}+i\eta_{\text{F}}$, where $\theta_{\text{F}}$ is the rotation of the major axis as compared to the incident linear polarization state, and $\eta_{\text{F}}$ is related to the ellipticity $e$ (the ratio of the minor and major axes) by $e=\tanh{\eta_{\text{F}}}$. The ellipsometric parameters $\eta_{\text{F}}$ and $\theta_{\text{F}}$ determined by

\begin{equation}\label{faradayrotationellipticity}
\theta_{\rm F}(\omega)=\frac{1}{2}\,\text{Im}\ \ln\left(\frac{E_+}{E_-}\right)
\quad
\eta_{\rm F}(\omega)=\frac{1}{2}\,\text{Re}\ \ln\left(\frac{E_+}{E_-}\right),
\end{equation}
are shown in Fig.~\ref{fig2}. Since the sample is much thicker than the attenuation length $c/(\omega\,\mbox{Im}\,n_{\pm})$ we can approximate Eqs.~\ref{faradayrotationellipticity} to

\begin{equation}\label{faradayrotationellipticity2}
\theta_{\rm F}(\omega)\simeq\frac{\omega d}{2c}\,\text{Re}\left(n_+-n_-\right)
\quad
\eta_{\rm F}(\omega)\simeq-\frac{\omega d}{2c}\,\text{Im}\left(n_+-n_-\right),
\end{equation}
where $d$ is the sample thickness. The lower limit of 0.1 meV constitutes the diffraction limit below which no radiation passes through the 8.5 mm wide aperture. The fact that $\theta_{\text{F}}(\omega)$ and $\eta_{\text{F}}(\omega)$ depend on the {\em relative} transmission of LCP and RCP photons removes most diffraction effects from these spectra. The result is very intuitive and shows that for a field of e.g.\,\,-1.3\,T, the absorption is strongly peaked at 0.14\,meV for the RCP chirality, hence the transmitted light has LCP chirality. Reversing the sign of the magnetic field reverses the direction of the local moments inside the material, hence now LCP photons are resonantly absorbed, allowing only RCP photons to pass through the material. This selective absorption of one particular chirality gives rise to a large Faraday rotation in a broad band of GHz radiation (Fig.~\ref{fig2}a) of more than 250\,\,deg\,mm$^{-1}$T$^{-1}$ which peaks up to an unprecedented level of 590 deg mm$^{-1}$ at 0.14\,meV (34\,GHz) at a relatively modest field of 1.6\,T.
The largest actually measured Faraday angle we obtained is about 560~deg at 1.6\,T for an optical path length of 0.945 mm (cf.~first echo in Fig.~\ref{fig1}). This rotation is well beyond measured angles in previously known record bulk materials: ferrites and garnets have typical rotations of 10 deg/mm in the microwave X band (around 0.04 meV). Mn$_{12}$Ac shows a rotation in a very narrow photon energy region around 1.25 meV of at most 130 deg (for a 0.75 mm thick sample) which is present at zero applied magnetic field due to the anisotropy field~\cite{vanslageren05}. The measured angle in the EuTiO$_3$ samples is also large as compared to rotations observed in thin materials like graphene and HgTe thin films where a Faraday rotation is induced by cyclotron resonance of charge carriers~\cite{crassee2011, shuvaev11}.
The weak shoulder around 0.25\,meV (cf.~Fig.~\ref{fig2}b) is due to wavelength dependent transmission through the 8.5 mm aperture in which the sample is mounted. Since the sample impedance for LCP and RCP depends on the magnetic field, this effect does not entirely cancel out in the ratio $E_+/E_-$.

\begin{figure}
\centering
\includegraphics[width=\columnwidth]{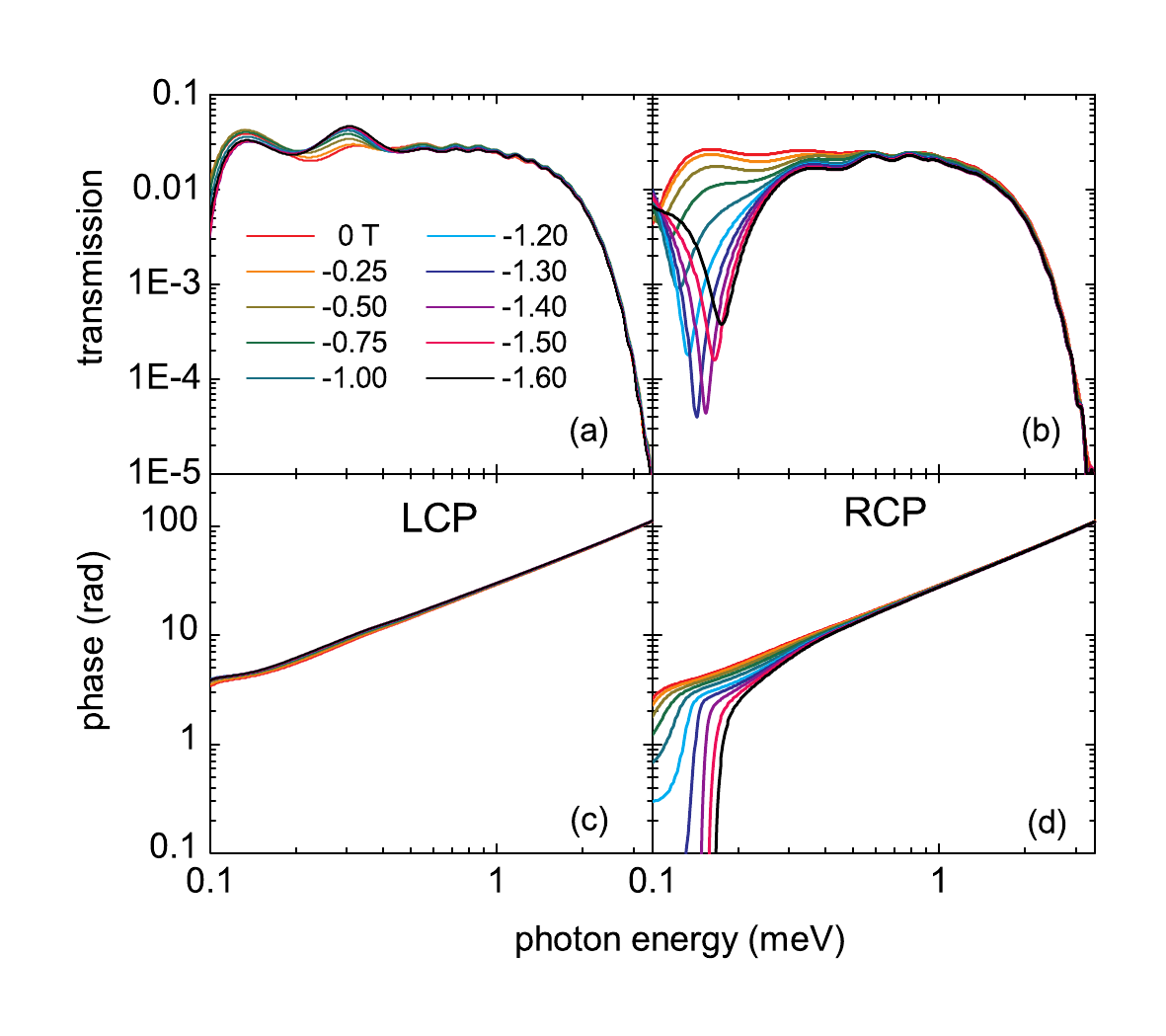}
\caption{Transmitted amplitude (a,b) and phase (c,d) of 265 $\mu$m thick EuTiO$_3$ as a function of photon energy for several external magnetic fields $H$ for LCP and RCP at 4.5\,K.}
\label{fig3}
\end{figure}

To obtain the transmission $T_\pm(\omega)$ of the first pulse, the signal transmitted through the sample was calibrated against a measurement of the empty aperture. The Fresnel equations $T_\pm(\omega)$ then are
\begin{equation}
T_\pm(\omega)=\frac{4 Z_\pm}{(1+Z_\pm)^2}\,e^{i(n_\pm-1)\omega d/c},
\label{straighttransmissioneq}
\end{equation}
where $Z_\pm=\sqrt{\epsilon_\pm/\mu_\pm}$. Fig.~\ref{fig3} shows the transmission amplitude and phase for LCP ($-$) and RCP (+) light between 0.1 meV and 3.5 meV at magnetic fields $H$ between 0\,T and $-1.6$\,T at 4.5 K. The curves show a pronounced absorption and a step-like phase increase for RCP light which strongly depend on the applied magnetic field whereas the transmission and phase for LCP light do not show any magnetic field dependence.

The observed circular dichroic phenomenon is present on an energy scale which corresponds to a magnetic dipole transition inside the Zeeman split $S=7/2$ moment of the Eu$^{2+}$ moments. Since in this case $g=2$, the Zeeman energy\index{Zeeman energy} is 0.12 meV per Tesla which is compatible with the observed resonance frequencies. If the resonance seen in the transmission spectra would originate from a peak in Im\,$\epsilon_{\pm}(\omega)$, this would imply huge changes of $\epsilon_{\pm}(0)$ as a function of $H$. However, this is excluded by the observation of Katsufuji \textit{et al.}~\cite{katsufuji01} where $\epsilon(0)$ only changes 7\,\% in a magnetic field of $1.6$\,T. Therefore the resonance seen in Fig.~\ref{fig3} must have its counterpart in the frequency dependence of $\mu_{\pm}(\omega)$.

\begin{figure}
\hspace{0mm}\includegraphics[width=\columnwidth]{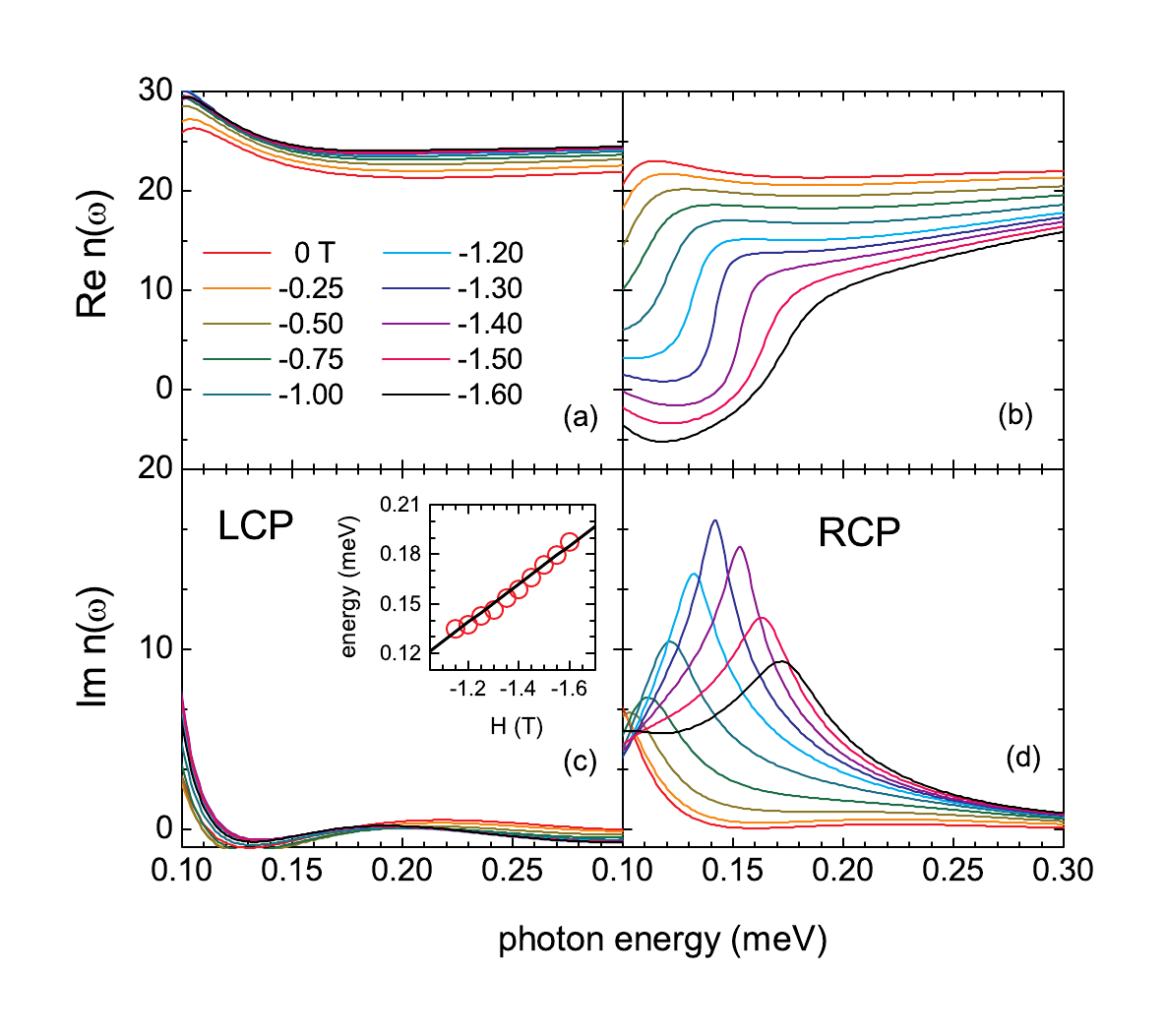}
\caption{Real and imaginary part of the index of refraction $n^\pm(\omega)=n_1^\pm(\omega)+i n_2^\pm(\omega)$, where $n(\omega)=\sqrt{\epsilon(\omega)\mu(\omega)}$, as a function of photon energy and magnetic field for LCP and RCP. The inset in (c) shows the energy of the maxima in $\mu_2^+(\omega)$ as a function of magnetic field (circles) and the Larmor energy $\hbar\omega_0(H)=0.1158H$ (black line).}
\label{fig4}
\end{figure}

In order to obtain $n_\pm(\omega)$ from the $T_{\pm}$ spectra we assume that far above the Zeeman energy  $\mu_\pm(\omega)\approx 1$. We used a single Lorentz oscillator corresponding to the ferroelectric soft mode to model $\epsilon_\pm(\omega)$. The parameters of this oscillator were fitted to the real and imaginary part of the transmission spectra between 1.25 meV and 3.7 meV. We finally calculated $n_\pm(\omega)$ between 0.1 and 3.5 meV by inversion of Eq.~\ref{straighttransmissioneq} using the low energy extrapolation of aforementioned parametrization of $\epsilon_\pm(\omega)\approx500$ (Fig.~\ref{fig4}). At energies above the resonance, Fig.~\ref{fig4} confirms that $n_2\rightarrow 0$ for both chiralities as expected. We verified that the real and imaginary part of $n_\pm(\omega)$ are Kramers-Kronig consistent, as required by causality.

At low energies the imaginary part  of $n_+(\omega)$ shows a peak which corresponds to the absorption of RCP light (cf.~Fig.~\ref{fig3}b). In order to verify the hypothesis that this absorption is due to purely magnetic dipole transitions within the Zeeman split Eu 4f levels, corresponding to a spin resonance, we plot the maxima of $\mu^+_2(\omega)$ as a function of the applied magnetic field $H$ (inset Fig.~\ref{fig4}c). This shows a linear behavior which goes through the origin when extrapolated to $H=0$. The figure also shows the Larmor energy $\hbar\omega_0=\hbar\gamma H$, with $\gamma=g\mu_{\text{B}}/\hbar$ for $g=2$, with $\mu_{\text{B}}$ the Bohr magneton. This is the energy required to bring a spin-system into resonance in the FM state.\cite{LaxButton} The perfect agreement illustrates the absence of an anisotropy field and thus of an orbital component in the magnetic ground state of the Eu ions. The width of the peak has the smallest value for an applied field of 1.4 Tesla, and may indicate a cross-over in the magnetic phase diagram characterized by a small value of the Gilbert damping~\cite{gilbert55}. The resonance contributes $\sim$0.1 to $\mu(\omega)$ for $\omega<\omega_0$, which is comparable to spin resonances observed in ferrite \cite{rado50}.

Interestingly the resonance shown in Fig.~\ref{fig4}d is a rather symmetric function of frequency for Im\,$n(\omega)$. Correspondingly, the real part of $n(\omega)$ has an almost purely dispersive line shape as expected from Kramers-Kronig relations between these two quantities. Yet, this observation is at first glance surprising, in view of the fact that we expect a symmetric resonance in Im\,$\mu(\omega)$ with dispersive counterpart in Re\,$\mu(\omega)$. [Note that $\epsilon(\omega)$ should be almost constant at these low frequencies.] Since $n(\omega)=\sqrt{\epsilon(\omega)\mu(\omega)}$, an asymmetric line shape should occur both in Re\,$n(\omega)$ and Im\,$n(\omega)$. Vice versa, the imaginary part of $n(\omega)^2$ (not shown) is strongly asymmetric. Several reasons for this behavior can be considered. $(i)$ In view of anticipated multiferroic properties of this material~\cite{fennie06} it is tempting to speculate that the asymmetry is induced by additional resonances at the same frequency occurring in $\epsilon(\omega)$ and/or the magneto-electric susceptibility. However, no mechanisms are known which can cause such strong coupling between spin and electric field. $(ii)$ Asymmetry could in principle arise from a departure of the condition of local response of  $n(k,\omega)$, in other words, if $n(k,\omega)$ depends strongly on the wave vector $k$. However, presumably the electric or magnetic response to the external field is confined to regions of the order of the grain-size of the polycrystalline sample (of order 10$^{-6}$~m), i.e., far smaller than the sample thickness thus excluding this scenario. $(iii)$ The wavefront has a distortion due to the 8.5 mm aperture. The wavefront distortion, and the corresponding transmission and phase, are necessarily wavelength dependent which becomes progressively weakened for wavelengths much smaller than the aperture. Since the spectral range of Fig.~\ref{fig4} corresponds to wavelengths between 4 and 12 mm, the third scenario offers the most plausible explanation for the observed line shape anomaly.

This work presents the first frequency and magnetic field dependence of the spin resonance in EuTiO$_3$. In contrast to other magnetic materials where ferromagnetic resonances have been observed~\cite{rado50,langner09} an orbital component is absent in the magnetic ground state of the Eu$^{2+}$ moments. We probe the transmission of EuTiO$_3$ by a novel spectroscopic technique of GHz time-domain ellipsometry, which provides the complex index of refraction function for right and left handed circular polarized light. These functions show a strong dichroism which causes the rotation of linearly polarized GHz light by a record amount of 590 deg/mm at 1.6 Tesla.

This work was supported by the Swiss National Science Foundation through through grant number 200020-125248 and the NCCR Materials with Novel Electronic Properties (MaNEP). T.K. was supported by GASR C-21560025 MEXT, Japan. We gratefully acknowledge stimulating discussions with J.N. Hancock and R. Viennois.

\end{document}